# Visualizing source code in 3D Maya software

**Ahmad Al- Shamailh1**

**1 Mutah University, Jordan**

**Ahmadsham013@gmail.com**

## ABSTRACT

*In this paper is clarify the summaries codes for programmers through three-dimensional shapes, and clearly programmers and developers, scholars and researchers in the field of software engineering, as well as researchers from the representative about three-dimensional forms. Through a three-dimensional drawing on a Maya scripts which are based on drawing shapes and three-dimensional stereoscopic show every part of the code, for example, classes, methods, coherence and homogeneity , In these drawings and show clearly and useful.*

## 1. Introduction

The rapid comprehensive development in the industrial fields, especially in recent years, has led to an increase in the need for a information technology and how to make it must be the existence of a computer as a tool to help in this technology. In the field of graphic design and be a key element. Also known within the field of computer-assisted drawing there is two dimensional systems 2D. The problem lies in the lack of understanding of the regulations and code to the rest of the team members understand the difficulty to reach a solution or a common and effective application to get to the good solutions to enable the understanding and awareness of the composition of the idea and its application.

The open nature and accessible, and researchers often analyze and audit these versions and history records system rather than processing all revisions of each file in the software system. The study indicates that the corrective actions, such as bug fixes, commits generate smaller than any other activity, the study of the statistical distribution in terms of size.

Binary systems configured boundaries of geometric means through several bilateral watershed dimensions that include details and important information about the drawing. In describing geometric shapes authority physical models, it is fully clear and using triple systems that are classified according to representation techniques, some of which describes the confusion by representation (edge-models), including what is described by (surface-models) and the other by describing the sizes of the forms. In addition, parts of the tripartite analysis and extract the necessary manufacturing data. And assist in the formation of good and clear representation of all members of the team or any user of the system to make it easier to understand and easy manner and expressed a clear, effective and simplified.

In these paper three-dimensional forms of the Maya program to produce stereoscopic fees help to extract the data and the conclusion is clear and simple way enjoyable through describing the Island of trees with plantings and each of them form certain significance. This work will be shown how to represent all of the code in detail and clarify parts of smoothing relations and interdependence between the elements of the code in general and method representation and the number of the calling between all method and the other and the extent of homogeneity and interdependence, remaining part of this paper is organizational part as so as later.

## 2. RELATED WORK

Normalization is more complex specifically if the number of relations and number of attributes in each relation is high [1]. Normalization when carried out manually can be time consuming, prone to errors and costly, since it needs high skilled personnel [2]. Thus, automating the process of normalization is the only solution for eliminate the drawbacks of manual normalization [3] [4].

Record date versions systems, such as CSV or vandalism, stores and record every change made in the warehouse, it represents a rich amount of information learned about the changes in the software system [5]. Understanding of the program is the main developer activity during software development and work on them, which represents more than half of the time spent on software maintenance and inspection. The developers face every day with software systems with thousands or millions of lines of code. Before try any changes to these systems, developers must understand parts of it. In addition to the note option is something in the middle, any offering developers a description of the source code [6]. Which can be read quickly and lead to a better understanding a city alone, and also permission to work a digital copy or paper of all or are awarded part of this work for personal or classroom without charge, have been used technology summarize and widely used in text analysis and is applied widely today by Internet search engines.

One of the major problems in software evolution is coping with the complexity, which stems from the huge amount of data that must be considered. A technique which can be used to reduce complexity in software visualization, good visual display allows human brain to study multiple aspects of complex problems parallel [7]. This research shows that it is classified matrices and development on the basis of representation in the form of layers and then we debate, case studies and information on the detail.

Remote sensing and limited hardware resources and capabilities in particular, after the development of software for these devices is certainly just as challenging for them as for other common computing platforms, and must be dealt with issues eventually coordination between the entire sensor network as a single unit. It is mainly composed parallel and great compatibility, distributed computational system. The programs running on actual hardware, it is almost completely opaque because of limited platform and a few channels I / O on the communication with the device. Also create usually a form of a model of the expected environment , so while the simulation is going to be very useful tools for the development of sensors, in the field of monitoring software can provide capabilities that help the developer a better understanding of the system they create or deploy added, and ultimately help to improve it and make sure it works well. We have been developing monitoring tool lightweight wireless sensor software [8].

Software systems are usually very large and complex for the average person to fully understand and properly. Software product metrics attempt to alleviate the problem by assembling and draw the details in order to expose the salient characteristics of the programs and understand them better. There is a need for future work to make the standards more useful, and there is a need for new tools to make measuring easier and more accurate. Is described EVOJAVA, a new tool is based on the source to measure the software structure of Java static code analysis, a particular goal of EVOJAVA is the ability to preserve the identity of the semantic features of the program, does not actually lead to improved software quality metrics[9]. The inclusions of time in our models also enable a family of metrics that measure the amount of aspects of software development, such as the age of the features or natural change rate. This standards development will help to answer questions such as whether the software quality metrics can be used to predict the likelihood of that maintenance programs at a later time. Information overload is already a challenge for software engineers and poses many difficulties, and the inclusion of another dimension of measurement data exacerbates the problem. Visualization techniques and support software tool necessary to communicate effectively measurements for developers.

The software companies build programs that can adapt to various changes and different work environments and diverse represents an important and mandatory goal of necessity can be achieved through the development of software, and this is the reason why large investments were directed towards the provision of free software. Also knows the reliability of the software and the possibility of free process a computer programmer in a specific environment failure for a specified period of time [10]. Many of the techniques of software to assist in the development of a test program before its release for public use and the great. Most of these techniques are just looking to build a prediction software that should have the ability to predict errors in the future under different test conditions models, and also the model is one of the successful models used in the literature to deal with dynamical systems .Comparison between AR and model known and rolling force model.

One of the possible scenarios for the classification of the program is a way to distinguish between 2D and 3D. Has been widely and successfully explore 2D visualization in the past, but There was no research on the detail enough in the visualization program on this side. 3D visualizations of existing programs, Should be need to enhance the 3D visualization through the loan firm and clear, as we discuss in section on related work, and this idea is not new, but the claim that until now has been used in the wrong levels of detail [11]. Choose the appropriate level of granularity is critical to support the city metaphor correctly. Categories are the essential foundation for the model and object-oriented, and along with the packages reside in, and this points the initial orientation of the developers. 3D visualization approach which creates cities that look real, because of a combination of designs, Topographic, and metric appointments applied to the appropriate level of granularity.

## 3. Switch source code to a three-dimensional drawing

In the beginning developers clarify the steps to work through the draw use case diagram to clarify the action steps in terms of inputs, processing and directed smart three-dimensional. In the future design tool to be inserted source code to this and turn it into a graphic three-dimensional but will be how to do it in this paper illustrate each specific code and how to turn them, and will be circulated this idea in the future on the codes big on large-scale projects under the umbrella engineering**,** Figure 1 shows.

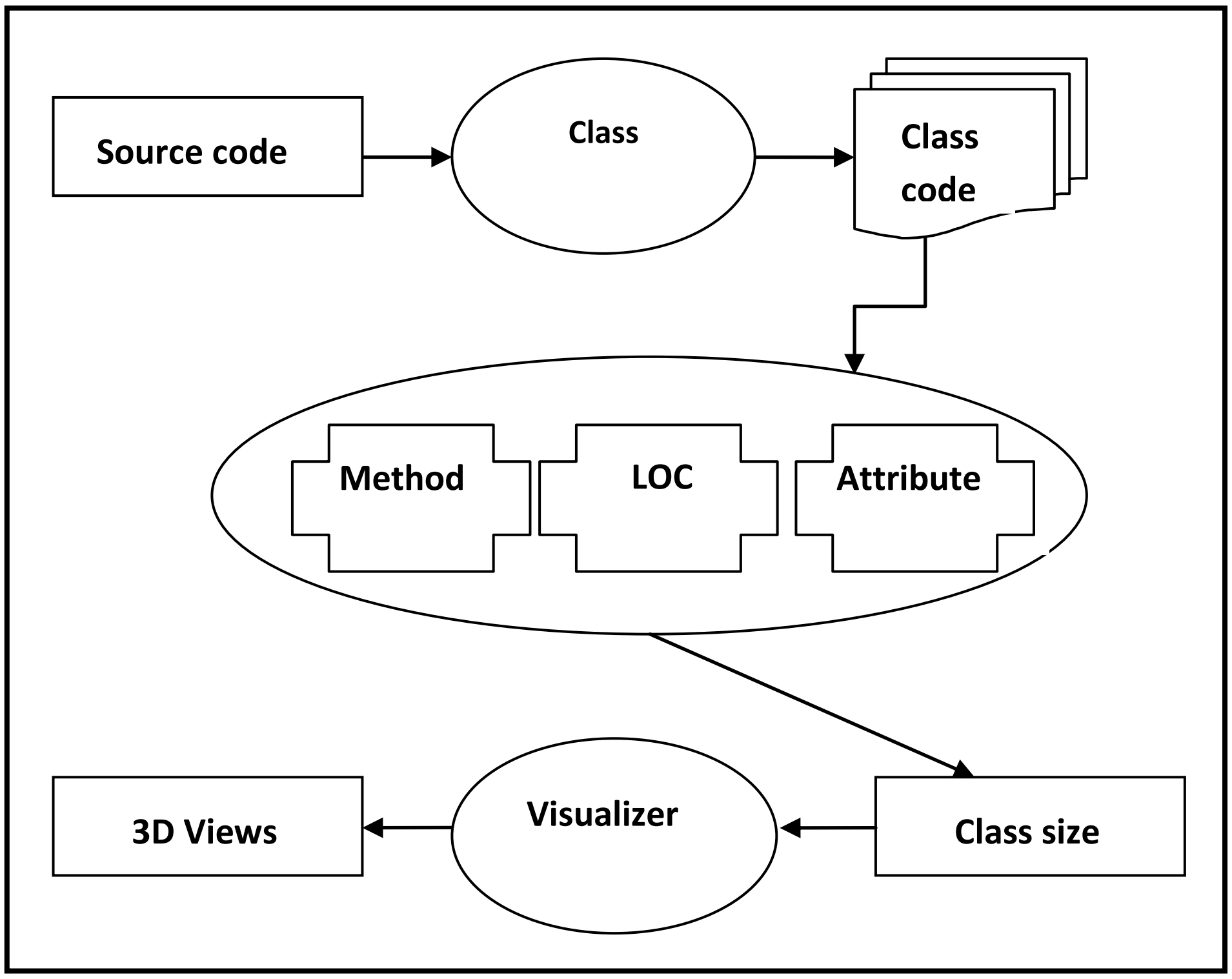

**Figure 1. Processes scheme**

Explained it distributed to determine the shapes on the corresponding code in who writes by programmers in software engineering, originally called. I will mention these models and corresponding in source code button on the form of garden and three-dimensional stereoscopic according to the following.

Each class in the project is represented as a tree and three-dimensional. As a result, number of these trees represents the number of the class in the project. And also, each level in the tree is set to represent method in the class. In each tree number method that contains it all class alone, as well as he sets the height of each tree on the number of lines in each class then represent the correlation between these trees in the form of links to water hyphen between the class knit and also fruits appear in every tree over the homogeneity between parts of the tree and the extent of heterogeneity between the types of methods for each class. In addition he gets this homogeneity winning number for each call method specific, and be representative of genetics during the arrival of spring water to all the class the children from the top to the bottom in the clear representation.

Here visualization examples convert code to three-dimensional drawing and applied to this example. In this example will be to clarify each class and number methods own three-dimensional drawing and each tree represents a number of the class and all methods number of leaves in our example the class binary number. In addition to the number of securities in both the two trees five cards in both, Figure 2 and Figure 3 show the three-dimensional painting for this example.

## Source code on example:

```
public class Useraaa {
  private String email;
  public String getEmail()
  {
      return email;
  }
  public void
  setEmail(String e){
      email = e;
  }
  public void notify(String
  msg) {
      // ....
  }
}
```

```
public class Ownerbbb
  extends Useraaa {
  private int
  maxNumLeagues;
  public int
  getMaxNumLeagues() {
      return maxNumLeagues;
  }
  public void
  setMaxNumLeagues(int n) {
      maxNumLeagues = n;
  }
}
```

**Figure 2 .Source code**

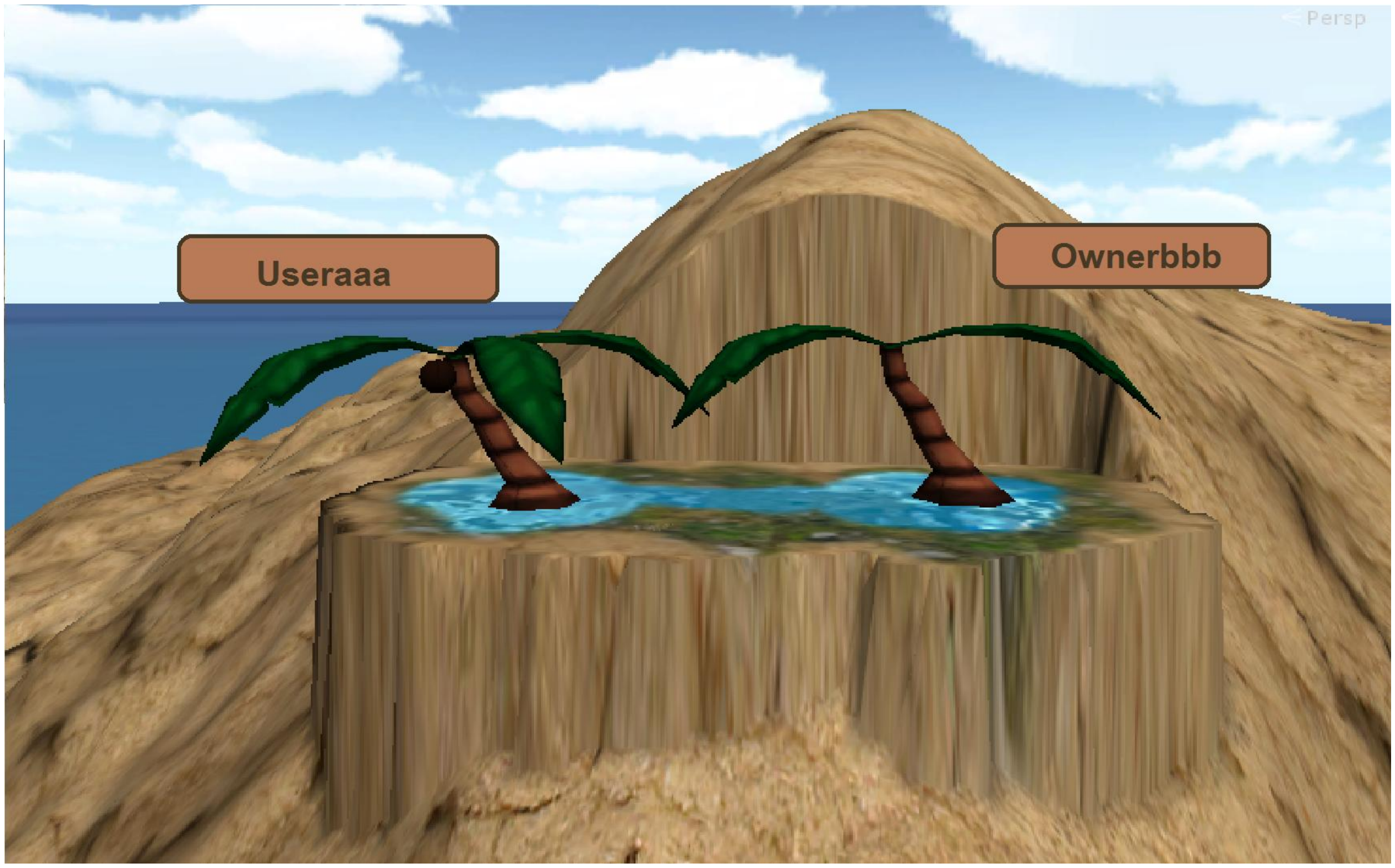

**Figure 3. Conversion code to representation**

## 4. Examples

In this example, the explain the idea of representation supported by details in pictures and here give detail each point previously mentioned and this example represents a the package "org.jfree.chart.annotations" taken from the following location "http://www.jfree.org/jfreechart/api/gjdoc/index.html" which contains fifteen Class and also a hundred and twenty-two method and the number of lines of code up to four thousand eight hundred and six ninety number and inherit up to twelve relationship and the extent of heterogeneity in each up to eleven relationship Inheritance here in the form of island contains what previously reported for easily understandable to the developers team. This shows in Figure 4.

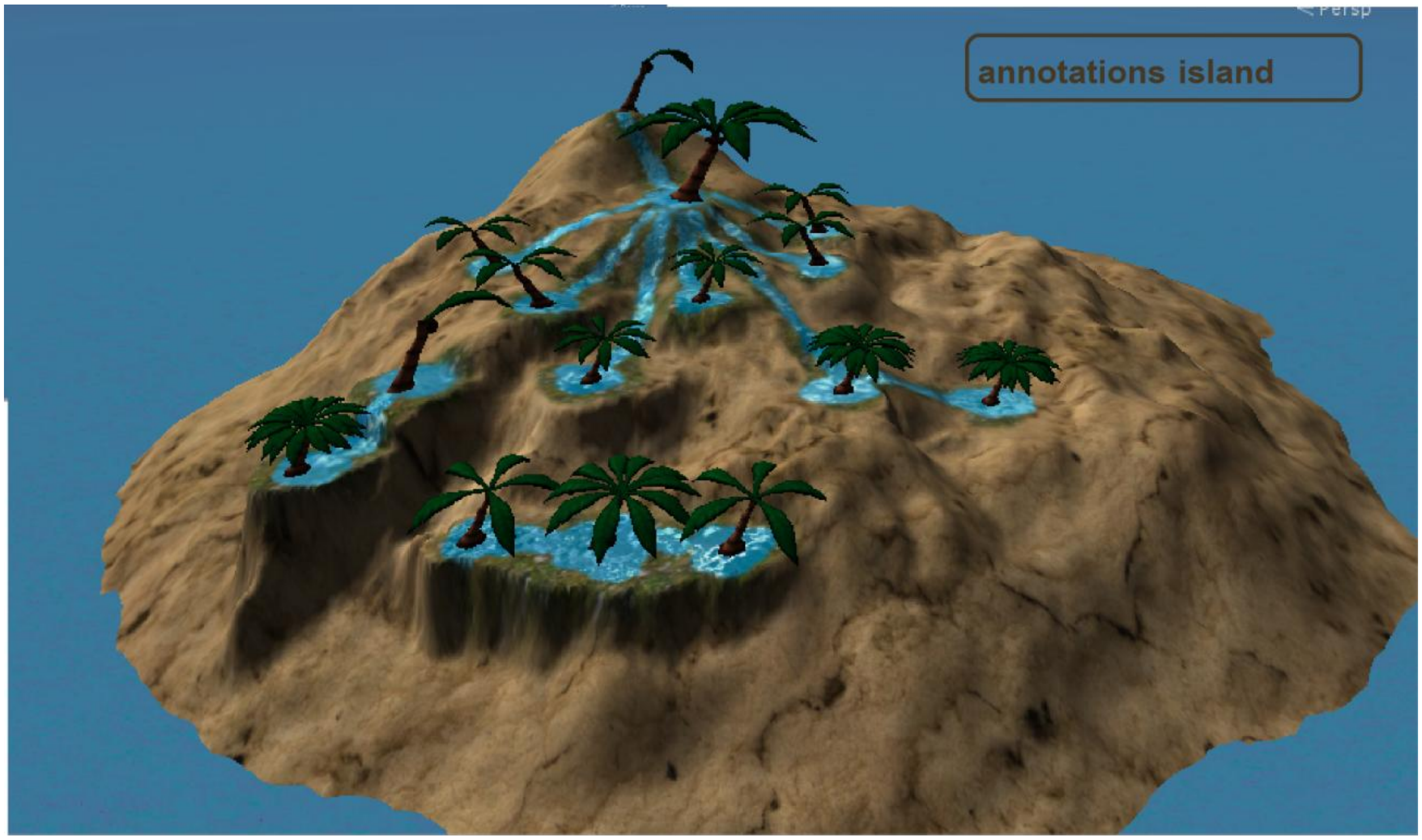

Figure 4. Example represents a the package annotations

Given the importance of the class names mentioned on the order of appearance in the source code as follows" CategoryAnnotation, XYAnnotation, AbstractXYAnnotation, CategoryLineAnnotation,CategoryPointerAnnotation,CategoryTextAnnotation, TextAnnotation,XYBoxAnnotation,XYDrawableAnnotation,XYImageAnnotation, XYLineAnnotation, XYPointerAnnotation, XYPolygonAnnotation, XYShapeAnnotation and XYTextAnnotation".

## 5. Determinants

It determinants that may encounter in this area that the conversion on the quantity and the momentum of the existing information in the statistics and data that may be available in the human discrimination of coherence and homogeneity may affect, this may lead to a shortage of learned information which may negatively affect the results of the conversion of graphics three-dimensional. This representation also needs a good skill in order to clarify the meaning of conversion code to a three-dimensional drawing properly and useful.

## 6. Conclusion

The result is that he is getting the forms of three-dimensional models of clear and logical help to extract the data and statistics, results and to provide support and service for both programmers and developers in software engineering or any science needs to be a three-dimensional representation later. In the future to serve a large segment of people to make it easier for them to understand they need to be aware of any use of these formats.